# Impact of lithium composition on the thermoelectric properties of the layered cobalt oxide system $Li_xCoO_2$

T. Motohashi,[1] Y. Sugimoto,[1] Y. Masubuchi,[1] T. Sasagawa,[2] W. Koshibae,[3] T. Tohyama,[4] H. Yamauchi,[2,5] and S. Kikkawa[1]

[1]*Faculty of Engineering, Hokkaido University, Sapporo 060-8628, Japan*
[2]*Materials and Structures Laboratory, Tokyo Institute of Technology, Yokohama 226-8503, Japan*
[3]*Cross-Correlated Materials Research Group (CMRG), RIKEN, Saitama 351-0198, Japan*
[4]*Yukawa Institute for Theoretical Physics, Kyoto University, Kyoto 606-8502, Japan*
[5]*Laboratory of Inorganic Chemistry, Department of Chemistry, Aalto University, FI-00076 AALTO, Finland*

Thermoelectric properties of the layered cobalt oxide system $Li_xCoO_2$ were investigated in a wide range of Li composition, $0.98 \geq x \geq 0.35$. Single-phase bulk samples of $Li_xCoO_2$ were successfully obtained through electrochemical deintercalation of Li from the pristine $LiCoO_2$ phase. While $Li_xCoO_2$ with $x \geq 0.94$ is semiconductive, the highly Li-deficient phase ($0.75 \geq x \geq 0.35$) exhibits metallic conductivity. The magnitude of Seebeck coefficient at 293 K ($S_{293K}$) significantly depends on the Li content ($x$). The $S_{293K}$ value is as large as +70 ~ +100 μV/K for $x \geq 0.94$, and it rapidly decreases from +90 μV/K to +10 μV/K as $x$ is lowered within a Li composition range of $0.75 \geq x \geq 0.50$. This behavior is in sharp contrast to the results of $x \leq 0.40$ for which the $S_{293K}$ value is small and independent of $x$ (+10 μV/K), indicating that a discontinuous change in the thermoelectric characteristics takes place at $x = 0.40$ ~ 0.50. The unusually large Seebeck coefficient and metallic conductivity are found to coexist in a narrow range of Li composition at about $x = 0.75$. The coexistence, which leads to an enhanced thermoelectric power factor, may be attributed to unusual electronic structure of the two-dimensional $CoO_2$ block.

# I. INTRODUCTION

Since the discovery of high temperature superconductivity in copper oxides, there have been intense interests in transition metal oxides with unconventional electronic behaviors. Among them, the layered cobalt oxide $Na_xCoO_2$ has been highlighted for its excellent thermoelectric properties [1]. It was revealed that $Na_xCoO_2$ exhibits unusually large thermoelectric power and metallic conductivity simultaneously, making this compound a promising thermoelectric material with high values of thermoelectric power factor $P \equiv S^2/\rho$ and figure-of-merit $Z \equiv S^2/\rho\kappa$ ($S$, $\rho$, and $\kappa$ are Seebeck coefficient, electrical resistivity, and thermal conductivity, respectively). The origin of the enhanced thermoelectric power factor was theoretically investigated based on one-electron band picture [2,3] or multi-electron picture involving electron correlations [4,5]. Nevertheless, the mechanism underlying the large thermoelectric power has not been fully understood.

The crystal structure of $Na_xCoO_2$ consists of alternate stacking of single-atomic Na layers and two-dimensional $CoO_2$ layers, and $Na_xCoO_2$ is known to show a wide range of Na nonstoichiometry [6]. Previous studies on $Na_xCoO_2$ evidenced a complex electronic phase diagram, which strongly depends on the Na content ($x$), including a spin-density-wave state at $x = 0.75$ [7], a charge-ordered state of poor electrical conduction at $x = 0.50$ [8], and superconductivity at $x \approx 0.35$ in a hydrated form [9]. These findings indicate that the formal cobalt valence, denoted as $V_{Co} = 4-x$, is one of the key parameters for the electronic properties of this oxide system. Despite the extensive research in previous works, there still remains a critical issue regarding the role of cobalt valence on the electronic properties. That is, it has been unclear whether the behaviors of $Na_xCoO_2$ are common features for the cobalt oxide family containing the two-dimensional $CoO_2$ layers. In fact, Na ions in $Na_xCoO_2$ tend to form superstructures with characteristic Na compositions [10-13]. It is suggested that the formation of superstructures obscures the intrinsic nature in the $CoO_2$ layer, as a Coulomb potential due to long-range Na-ion order may influence the adjacent $CoO_2$ layer.

To address this issue, we focused on the layered cobalt oxide, $Li_xCoO_2$. This compound is regarded as a potential reference material of $Na_xCoO_2$, as both $Li_xCoO_2$ and $Na_xCoO_2$ contain the $CoO_2$ layer in common. It is widely recognized that the capability of Li intercalation/deintercalation is remarkable for $Li_xCoO_2$ such that Li nonstoichiometry can be precisely controlled through electrochemical technique [14,15]. Note that this oxide is one of the representative cathode materials for the rechargeable Li-ion battery. Moreover, $Li_xCoO_2$ exhibits essentially an identical stacking pattern in a wide range of Li composition, $1.0 \geq x \geq 0.25$, without forming long-range Li-ion order except for $x = 0.50$ [16]. Recently, we successfully obtained a sample series of single-phase $Li_xCoO_2$ through electrochemical deintercalation of Li from pure $LiCoO_2$ ($x = 1.0$) bulks [17,18]. Based on dc-magnetic-susceptibility data, combined with results of $^{59}Co$ nuclear magnetic resonance (NMR) and nuclear quadrupole resonance (NQR) observations [19], the electronic phase diagram of $Li_xCoO_2$ was established [20]. The phase diagram of $Li_xCoO_2$ is complex and the electronic properties are sensitively influenced by the Li content ($x$). Here, we report measurements of electrical resistivity ($\rho$) and Seebeck coefficient ($S$) of $Li_xCoO_2$ with Li contents of $0.98 \geq x \geq 0.35$. The thermoelectric properties of $Li_xCoO_2$ are discussed in terms of the Li composition ($x$) and thereby the cobalt valence ($V_{Co}$). The present result provides new insights into the electronic structure of the cobalt oxide thermoelectrics and thus contributes to the comprehensive understanding of the unusually large thermoelectric power.

## II. EXPERIMENT

The $Li_xCoO_2$ samples used in the thermoelectric measurements were synthesized through electrochemical deintercalation of Li from pristine $LiCoO_2$, as described elsewhere [17,18,20]. Approximately 50 mg of polycrystalline $LiCoO_2$ pellet was electrochemically oxidized with a constant current of 0.05 − 0.1 mA, i.e., the galvanometric setup. The $LiCoO_2$ pellet (cathode) and an aluminum metal disk (anode) were set in an airtight flat cell filled with a non-aqueous electrolyte. No auxiliary agents (i.e., carbon black and Teflon powder) were added to the bulk pellets to avoid extrinsic contributions. For each sample, the Li content (or the amount of Li ions to be extracted, i.e., 1–$x$) was precisely controlled by the reaction duration based on Faraday's law with an assumption that the full amount of electricity due to the current was used for the electrochemical deintercalation of Li. Typically, 50 mg sample was charged with $I = 0.1$ mA for 34.2 h, 68.5 h, and 89.0 h to obtain the $x$ = 0.75, 0.50, and 0.35 phases, respectively.

After the electrochemical procedure, the $Li_xCoO_2$ samples were washed with anhydrous dimethyl carbonate and cut into rectangular pieces in an argon-filled glovebox. The sample sizes were typically 4 × 2 × 0.5 mm$^3$ and 2 × 3 × 0.2 mm$^3$ for electrical resistivity and thermoelectric power measurements, respectively. Since the high-valent cobalt oxides tend to experience chemical instability with atmospheric moisture, sample handling was made such that the samples might be exposed to moisture as little as possible (within several minutes). The X-ray powder diffraction (XRPD) analysis indicated that such short exposure to air does not cause deterioration and decomposition in our samples. Electrical resistivity ($\rho$) was measured employing a four-probe technique (Quantum Design; PPMS) in a temperature range of 2 − 300 K. Seebeck coefficient ($S$) was measured with a steady-state technique at temperatures between 5 and 293 K. The measurements were performed on several sample batches to check reproducibility. In addition, for some selected samples $\rho$ and $S$ were measured simultaneously to ensure that a set of the thermoelectric data is indeed obtained from a single Li composition. These measurements were carried out utilizing a home-made apparatus equipped with a helium closed-cycle refrigerator.

## III. RESULTS

### A. Sample preparation

Polycrystalline $Li_xCoO_2$ samples of $x$ = 0.98, 0.96, 0.94, 0.75, 0.67, 0.60, 0.50, 0.40, and 0.35 were synthesized. The XRPD analysis revealed that all the samples are of single phase with good crystallinity, indicating high degree of chemical homogeneity. For each sample, the diffraction peaks were readily indexed based on the structural model reported previously [16,21-24]. The lattice parameters of each sample were in good agreement with those in the literature. Details in the sample characterization such as chemical composition and crystal structure are given elsewhere [20].

It is widely recognized that at nominal Li contents 0.94 > $x$ > 0.75 single-phase products are never obtained but two $Li_xCoO_2$ phases with distinct Li contents coexist (i.e., the $x$ = 0.94 and 0.75 phases) [25-27]. Meanwhile, the chemical stability was found to significantly deteriorate for $x \leq 0.25$: diffraction peaks of the $x \leq 0.25$ samples became broadened immediately when exposed to atmospheric moisture. Taking into account these facts, we excluded samples with $0.94 > x > 0.75$ and $x \leq 0.25$ from the present investigation.

**B. Electrical resistivity**

Figures 1, 2, and 3 show the dependence of $\rho$ on temperature for the Li$_x$CoO$_2$ samples. It appears that the behaviors significantly depend on the Li content ($x$). Focusing on the absolute value and temperature variation of $\rho$, the samples are categorized into the following three groups depending on their Li compositions.

(1) *Higher Li concentrations: $x \geq 0.94$* (Fig. 1)—the samples in this regime exhibit relatively large $\rho$ values and semiconductive behaviors at low temperatures. The magnitude of electrical resistivity at 293 K ($\rho_{293K}$) are 0.8 and 0.5 $\Omega$ cm for $x = 0.98$ and 0.94, respectively. The $x = 0.98$ sample shows a pronounced upturn in the $\rho - T$ curve, while the upturn is significantly weakened when $x$ is decreased only by 0.04 (i.e., at $x = 0.94$).

(2) *Intermediate Li concentrations: $0.75 \geq x \geq 0.50$* (Fig. 2)—the samples in this regime are featured with metallic conductivity with small $\rho$ values. In fact, the $\rho$ value is typically $10^{-2} - 10^{-1}$ $\Omega$ cm at room temperature, being an order of magnitude smaller than that for $x \geq 0.94$. It is also noteworthy that all the samples exhibit anomalies in the $\rho - T$ curves. The temperatures at which the anomalies appear are closely related to the Li contents ($x$) and in good agreement with the magnetic anomaly points [20]. The $\rho - T$ curve of the $x = 0.50$ sample is most unusual: the $\rho$ value suddenly increases below $T_t = 120 - 170$ K involving temperature hysteresis of $\Delta T = 50$ K, which is much larger than those for $x = 0.60$, 0.67, and 0.75 ($\Delta T \approx 4$ K).

(3) *Lower Li concentrations: $x \leq 0.40$* (Fig. 3)— the samples in this regime also show metallic conductivity. The magnitude of $\rho$ is as large as $4 - 5 \times 10^{-2}$ Ω cm at room temperature: the value is comparable to that for the intermediate Li concentrations ($0.75 \geq x \geq 0.50$). Nevertheless, the temperature dependence of $\rho$ is clearly different. The $\rho - T$ curves are metallic in the whole temperature range below 300 K, without exhibiting any anomalies.

In Fig. 4, the $\rho_{293K}$ value is plotted as a function of the Li content ($x$), or the formal cobalt valence ($V_{Co}$). It can be seen that the magnitude of $\rho$ decreases greatly when the Li content is reduced beyond the two-phase region located at $0.94 \geq x \geq 0.75$. A similar behavior was also reported by Ménétrier *et al.* [27] (see blue triangles in Fig. 4). The previous work by Marianetti *et al.* suggested [28] that the two-phase coexistence in the Li$_x$CoO$_2$ system is the consequence of a first-order metal-nonmetal transition triggered by strong correlation effects of Co 3$d$ electrons. In the nonmetal phase with $x \geq 0.94$, Li vacancies may bind holes resulting in an impurity band. This band is half-filled because there is one hole per vacancy, and it will be split to form a Mott insulator due to the on-site Coulomb interaction. For $x \leq 0.75$, on the other hand, the holes become delocalized leading to a metallic state as the hole density is enough high to effectively screen the vacancy potential. As mentioned above, the metallic Li$_x$CoO$_2$ phase can be divided into two groups in terms of the temperature variation of $\rho$, i.e., $0.75 \geq x \geq 0.50$ and $x \leq 0.40$, whereas these groups look very similar to each other when focusing on the absolute value of $\rho_{293K}$.

### C. Seebeck coefficient

Figure 5 shows Seebeck coefficient ($S$) vs. temperature plots for some representative samples. The data demonstrate that the thermoelectric property is sensitively influenced by the Li content ($x$). For Li-rich compositions $x \geq 0.5$, the temperature variation is remarkable. The behavior is particularly complex for $x = 0.50$. Noticeable features of this sample are that upon cooling the sign of $S$ changes from positive to negative, and the magnitude of $S$ (with negative sign) abruptly increases below $T_t = 170$ K. The anomaly is likely to originate from an electronic phase transition involving charge ordering or charge disproportionation triggered by a fractional cobalt valence $V_{Co} = +3.50$ [20]. On the other hand, the $x \leq 0.40$ samples show nearly $T$-independent Seebeck coefficient at $T \geq 100$ K with small positive values typically seen in conventional metals. This means that a discontinuous change in the thermoelectric characteristics takes place at $x = 0.40 - 0.50$. It should be noted that the behavior of electrical resistivity also varies at the same Li composition, suggesting the existence of the critical point in the electronic phase diagram of Li$_x$CoO$_2$.

In Fig. 6, the magnitude of Seebeck coefficient at 293 K ($S_{293K}$) is plotted as a function of $x$ and $V_{Co}$: the strong compositional dependence is evident for $S_{293K}$. As we mentioned, samples with metallic conductivity can be divided into two groups: i.e., $x \leq 0.40$ and $0.75 \geq x \geq 0.50$. In the former group, $S_{293K}$ is small in magnitude and independent of $x$. In contrast, the $S_{293K}$ value of the latter group linearly increases as the Li content increases from +10 µV/K ($x = 0.50$) to +90 µV/K ($x = 0.75$), even though the $\rho_{293K}$ value is almost constant in the whole range of metallic compositions (see Fig. 4). Consequently, large thermoelectric power and small electrical resistivity appear simultaneously in the vicinity of $x = 0.75$. Another important feature is that the $S_{293K}$ values for $x \geq 0.94$ are comparable to that for $x = 0.75$, even though the $x \geq 0.94$ samples are nonmetallic with much larger $\rho$ values than the metallic $x = 0.75$ phase. Obviously, this behavior is in sharp contrast to the general trend of conventional semiconductors, in which both the $\rho$ and $S$ values are accordingly enhanced as the number of carriers decreases [29].

From the results of the $\rho$ and $S$ measurements, the thermoelectric power factor $P \equiv S^2/\rho$ was calculated (Fig. 7). The magnitude of $P$ is significantly enhanced at $x = 0.75$, i.e., the end member of metallic compositions, reflecting the coexistence of large $S$ and small $\rho$. Meanwhile, the $P$ value becomes smaller when the Li content increases or decreases, due to the larger $\rho$ or smaller $S$, respectively. It is thus revealed that the remarkable thermoelectric behavior appears *only* in a narrow range of Li composition at about $x = 0.75$. Apparently, the variation of Seebeck coefficient with respect to carrier concentration differs from that for conventional semiconductors, implying unusual carrier dynamics in Li$_x$CoO$_2$.

## IV. DISCUSSION

The present work has revealed a large impact of Li composition on the thermoelectric property of Li$_x$CoO$_2$. This oxide system is divided into three regimes in terms of electrical resistivity and Seebeck coefficient: i.e., the higher ($x \geq 0.94$), intermediate ($0.75 \geq x \geq 0.50$), and lower ($x \leq 0.40$) Li compositions. Marianetti *et al.* suggested [28] that Li$_x$CoO$_2$ undergoes a first-order Mott transition at $0.94 \geq x \geq 0.75$, which separates metallic and nonmetallic phases with lower/intermediate and higher $x$ values, respectively. The measurements of $\rho$ and $S$ in the present work have indicated that another boundary exists at $x = 0.40 - 0.50$. Importantly, the magnetic behavior is also found to vary at the critical Li content, $x_c = 0.35 - 0.40$: the magnetism looks like of Curie-Weiss type for larger $x$, while it is Pauli paramagnetic for lower $x$ [20]. The accordance of these critical points implies that a distinct change in the electronic structure takes place at the boundary. Among the three compositional regimes, the intermediate one ranging from $x = 0.50$ to $0.75$ is particularly worth of attention. *Only* in this regime, thermoelectric power is greatly enhanced with $x$, while the Li$_x$CoO$_2$ phase retains its metallic conductivity. The appearance of the Curie-Weiss type magnetism is somewhat anomalous from the viewpoint of a paramagnetic metal, suggesting the importance of electron correlation effects. It is thus likely that the strongly-correlated electronic structure plays a key role in the remarkable thermoelectric performance of Li$_x$CoO$_2$.

There are earlier works on the transport properties of Li$_x$CoO$_2$. Ménétrier *et al.* reported electrical resistivity and thermoelectric power of polycrystalline Li$_x$CoO$_2$ bulks prepared through electrochemical deintercalation of Li from LiCoO$_2$ [27]. The authors claimed that upon Li deintercalation a metal-nonmetal transition occurs at $0.94 \geq x \geq 0.75$ involving two-phase coexistence: this finding is in good agreement with our observations. Since their work mainly focused on the mechanism of the two-phase coexistence, properties of the lower Li compositions ($x < 0.50$) were not reported. On the other hand, Miyoshi *et al.* investigated electrical resistivity of Li$_x$CoO$_2$ single crystals in a wide range of Li compositions $0.25 \leq x \leq 0.99$ [30]. Also, Ishida *et al.* reported measurements of $\rho$ and $S$ on Li$_x$CoO$_2$ thin films with $0.66 \leq x \leq 0.87$ [31]. Although these works reached similar conclusions to ours concerning the evolution of metallic behaviors in the Li-deficient phase, some discrepancies are seen in the temperature variations of $\rho$ and $S$. In fact, signatures of the critical point at $x = 0.40 \sim 0.50$ were never discussed in these works, because distinct changes in the thermoelectric characteristics were hardly seen at this Li composition. It should be noted that their Li$_x$CoO$_2$ samples were prepared through a complicated route: the samples were obtained with an ion exchange reaction between Na$_x$CoO$_2$ and Li$_2$CO$_3$/LiNO$_3$, followed by Li extraction utilizing chemical reagents such as NO$_2$BF$_4$ and K$_2$S$_2$O$_8$ [30,31]. We suggest that the difference in the synthesis routes may affect the degree of chemical homogeneity, which would be the main source of the discrepancies in the experimental results.

It is important to note that some common features can be seen in the Li$_x$CoO$_2$ and Na$_x$CoO$_2$ systems. Indeed, the following experimental facts were reported for Na$_x$CoO$_2$, which are closely similar to those for Li$_x$CoO$_2$. (1) A magnetic critical point exists in the vicinity of $x_c \approx 0.5$, which separates the Pauli-paramagnetic and Curie-Weiss metals [8,32,33]. (2) The thermoelectric property significantly depends on the Na content [34-37]. The $S$ value is greatly enhanced as $x$ increases in the Curie-Weiss regime (i.e., $x > x_c$), while it is small with positive sign in the Pauli-paramagnetic regime ($x < x_c$). The electronic structure of Na$_x$CoO$_2$ was theoretically studied by Lee *et al.* on the basis of first principle calculations [38,39]. They concluded that the phase diagram of this oxide is characterized by a crossover from effective single-band character with strong electron correlations for $x > x_c$ into a multiband regime $x < x_c$, where correlation effects are substantially reduced. We emphasize that their picture is also consistent with our observations for $Li_xCoO_2$. The similarities in the electronic properties imply that the essential physics within the $CoO_2$ block is identical in both the $Li_xCoO_2$ and $Na_xCoO_2$ systems. The evolution of thermoelectric properties with varying cobalt valence ($V_{Co}$) may be the common feature in the layered cobalt oxide family.

Despite the similarities in the features of $Li_xCoO_2$ and $Na_xCoO_2$, a careful comparison of the $S - T$ data reveals some quantitative differences between the two systems. First, the magnitude of $S$ is somewhat smaller for $Li_xCoO_2$ than that for $Na_xCoO_2$ [34-36], e.g., $S_{293K} = 80 - 90$ μV/K and $100 - 120$ μV/K at $x = 0.75$ for the former and the latter, respectively. Second, in the higher $x$ regime the $S$ value of $Li_xCoO_2$ increases almost linearly with temperature, while the $S$ value of $Na_xCoO_2$ seems to approach a certain value at higher temperatures [34-37]. We interpret these differences as the consequence of weaker electron correlations of $Li_xCoO_2$ than $Na_xCoO_2$. In this context, it is worth noting that layered rhodium oxides $K_xRhO_2$ [40] and $Sr_{1-x}Rh_2O_4$ [41], which are isomorphous to $Na_xCoO_2$, exhibit smaller $S$ values with a nearly $T$-linear behavior in a wide range of temperature. This experimental fact has been explained on the basis of more itinerant nature of Rh $4d$ electrons than Co $3d$: in other words, the weaker electron correlations of Rh $4d$ are attributable to the characteristic behaviors of the rhodium oxides.

The origin of the weaker correlation effects in $Li_xCoO_2$ has not been well understood. We tentatively believe that the weaker electron correlation of $Li_xCoO_2$ is the consequence of the more three-dimensional electronic structure due to the shorter interlayer Co-Co distance in $Li_xCoO_2$ than $Na_xCoO_2$, i.e., $d_{Co-Co}$ = 4.7 – 4.8 Å and 5.4 – 5.5 Å for the former and the latter, respectively. There are earlier works of first principle calculations on $A_xCoO_2$ with $A$ = Li [42], Na [2,43], and K [43]. While all the $A_xCoO_2$ members have a common feature in the valence band $t_{2g}$ manifold, the direct comparison of the band structure is not possible between $Li_xCoO_2$ and the other members, as the fine structure near the Fermi level has never been discussed for $Li_xCoO_2$ by means of the LDA + U (local density approximation of Hubbard U) method. A detailed theoretical study is thus highly desirable for $Li_xCoO_2$ to get additional insights into the electronic structure.

Finally, we discuss the transport properties of semiconductive $Li_xCoO_2$ with higher Li concentrations. As presented in Figs. 1 and 5, the $x \geq 0.94$ samples show resistivity upturn and $T$-linear Seebeck coefficient. To ensure this feature, $\rho$ and $S$ of the $x$ = 0.98 sample were measured simultaneously, and it was indeed confirmed (Fig. 8). One may notice that this result is somewhat unusual within general aspects of conventional semiconductors. It is widely recognized [44] that electrical resistivity and thermoelectric power of semiconductors are closely related to each other, and the temperature dependences of $\rho$ and $S$ are given by $\rho \propto \exp(\Delta/T)$ and $|S| \propto \Delta/T$ ($\Delta$ denotes the energy gap of the semiconductor) under an assumption that the carriers are thermally activated according to $n \propto \exp(-\Delta/T)$. This argument leads to concomitant increases in $\rho$ and $S$ at low temperatures, apparently in contradiction to our observation. It should be emphasized that such "conflicting" behaviors were already reported by Ménétrier *et al.*

[27] who observed the $T$-linear Seebeck coefficient in their semiconductive samples (i.e., $x \geq 0.94$). The origin of the $T$-linear Seebeck coefficient in semiconductive Li$_x$CoO$_2$ has remained unclear. We suggest that this feature is significant and would be closely related to the unconventional thermoelectric behaviors of Li$_x$CoO$_2$.

## V. CONCLUSIONS

Single-phase samples of the layered cobalt oxide Li$_x$CoO$_2$ were prepared through electrochemical deintercalation of Li, and their electrical resistivity ($\rho$) and Seebeck coefficient ($S$) were measured in a wide range of Li composition, $0.98 \geq x \geq 0.35$. This oxide system is divided into three distinct regimes in terms of its thermoelectric behaviors, as summarized in Table 1. It appeared that *only* in a narrow range of Li composition at about $x = 0.75$, large thermoelectric power and metallic conductivity appear simultaneously, leading to the remarkable thermoelectric performance of Li$_x$CoO$_2$.

The thermoelectric behaviors are similar between the Li$_x$CoO$_2$ and Na$_x$CoO$_2$ systems. The similarities imply that the evolution of thermoelectric properties with varying Li/Na content ($x$) or formal cobalt valence ($V_{Co}$) may be the common feature in the layered cobalt oxide family. Based on the finding in the present work, a strategy to design thermoelectric cobalt oxides can be suggested: i.e., the marginally doped samples in the metallic regime tend to show small $\rho$ and large $S$ simultaneously, thereby an enhanced thermoelectric power factor. Such a "moderately-doped CoO$_2$ block" is a key factor to realize the excellent thermoelectric property which is likely originating from electron correlation effects.

## ACKNOWLEDGMENTS

The authors thank Prof. M. Karppinen (Aalto University) and Dr. M. Mori (Advanced Science Research Center, Japan Atomic Energy Agency) for their fruitful discussion and comments. Also, Prof. Y. Hinatsu and Prof. M. Wakeshima (Hokkaido University) are acknowledged for their help in the electrical resistivity measurements. The thermoelectric measurements were conducted in part through the Collaborative Research Project of Materials and Structures Laboratory, Tokyo Institute of Technology. The present work was supported by Grants-in-aid for Science Research (Contracts No. 16740194 and No. 19740201) from the Japan Society for the Promotion of Science. T.M. acknowledges financial support from the Murata Science Foundation, and H.Y. from the Tekes (Grant No. 1726/31/07).

TABLE 1.

Three compositional regimes in the Li$_x$CoO$_2$ system and their characteristic features.

| Li content ($x$) | $x \leq 0.40$ | $0.75 \geq x \geq 0.50$ | $x \geq 0.94$ |
|---|---|---|---|
| Electrical resistivity ($\rho$) | small | small | large |
| Seebeck coefficient ($S$) | small | large | large |
| Thermoelectric power factor ($P \equiv S^2/\rho$) | small | large | small |
| Magnetism | Pauli paramagnetic | Curie-Weiss type | Curie-Weiss type |
| Electron correlations | weak | strong | strong |

Figure captions

FIG. 1. (Color online)

Temperature dependence of electrical resistivity ($\rho$) for the Li$_x$CoO$_2$ samples with $x$ = 0.94 and 0.98.

FIG. 2. (Color online)

Temperature dependence of electrical resistivity ($\rho$) for the $x$ = 0.50, 0.60, 0.67, and 0.75 samples.

FIG. 3. (Color online)

Temperature dependence of electrical resistivity ($\rho$) for the $x$ = 0.35 and 0.40 samples.

FIG. 4. (Color online)

The magnitude of electrical resistivity at 293 K ($\rho_{293K}$) as a function of $x$ in Li$_x$CoO$_2$. For comparison, data plots of Ménétrier *et al.* (Ref. 27) are also shown with blue triangles.

FIG. 5. (Color online)

Temperature dependence of Seebeck coefficient ($S$) for the $x$ = 0.35, 0.40, 0.50, 0.60, 0.67, 0.75, and 0.94 samples.

FIG. 6. (Color online)

The magnitude of Seebeck coefficient at 293 K ($S_{293K}$) as a function of $x$ in Li$_x$CoO$_2$.

FIG. 7. (Color online)

The thermoelectric power factor $P \equiv S^2/\rho$ for the Li$_x$CoO$_2$ samples. The $P$ values were calculated from the $\rho$ and $S$ values at 293 K.

FIG. 8.

Electrical resistivity ($\rho$) and Seebeck coefficient ($S$) of the $x = 0.98$ sample. These measurements were carried out simultaneously with a single piece of the bulk sample.

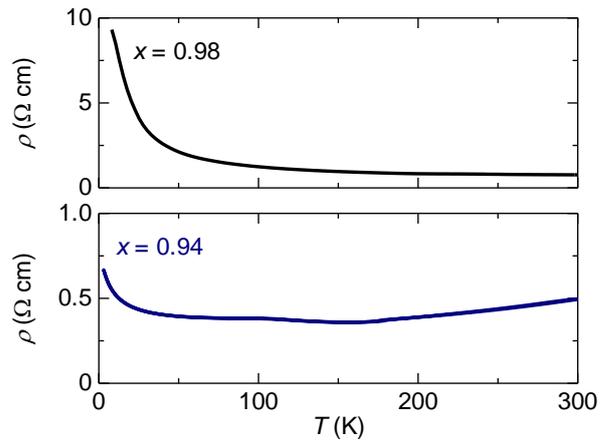

Fig. 1. Motohashi *et al.*

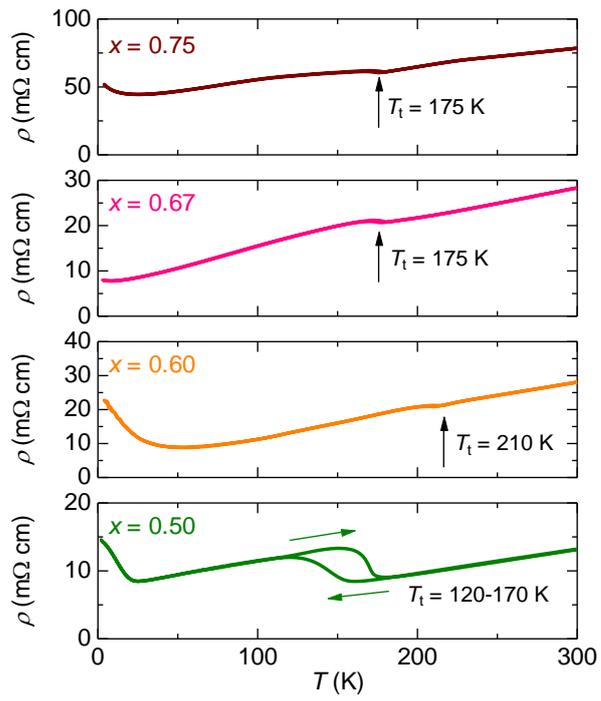

Fig. 2. Motohashi *et al.*

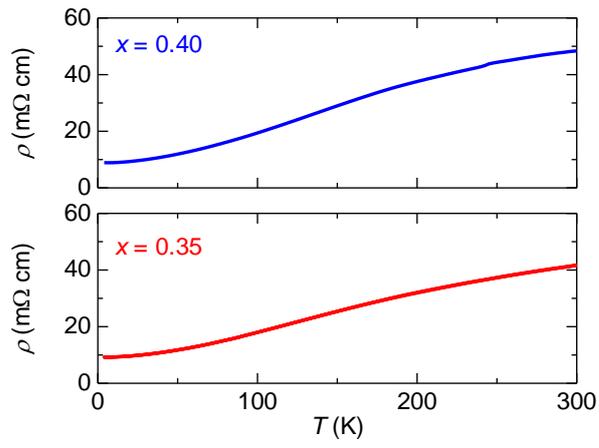

Fig. 3. Motohashi *et al.*

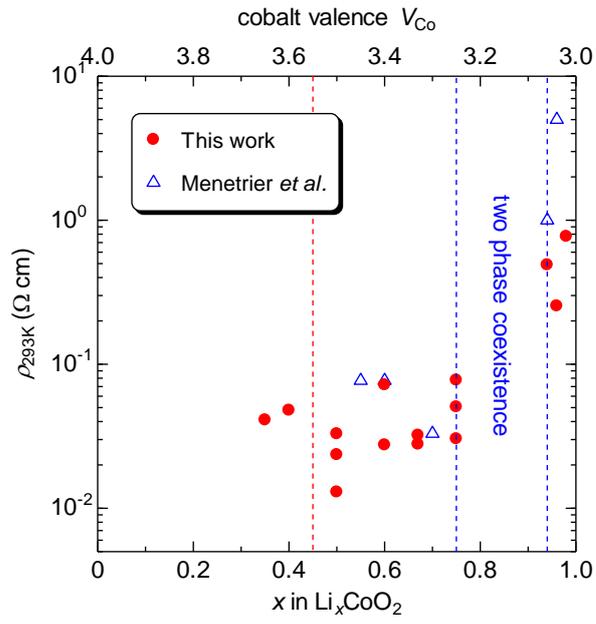

Fig. 4. Motohashi *et al.*

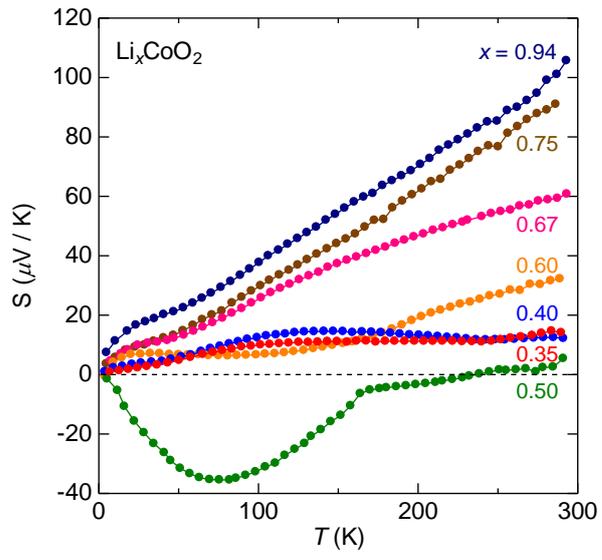

Fig. 5. Motohashi *et al.*

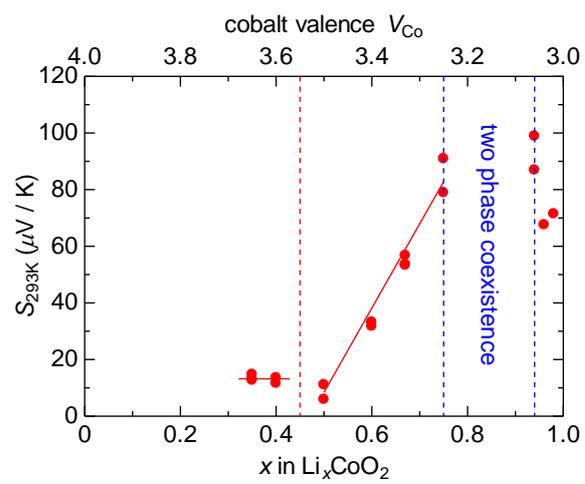

Fig. 6. Motohashi *et al.*

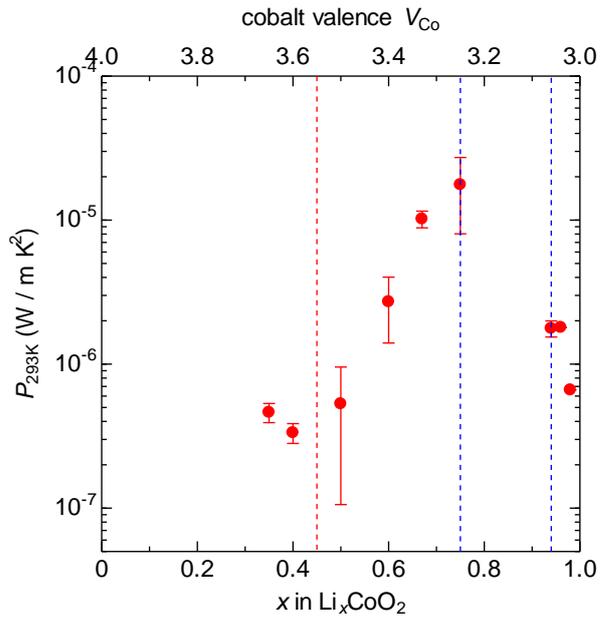

Fig. 7. Motohashi *et al.*

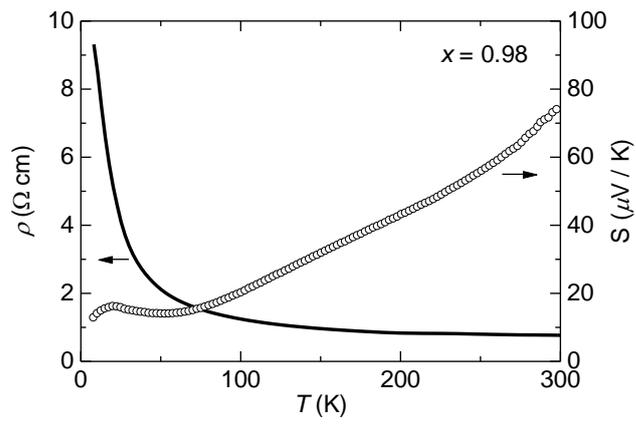

Fig. 8. Motohashi *et al.*